\documentclass[]{spie}
\usepackage{epsfig}
\usepackage{subfigure}
\usepackage{lscape}
\usepackage{graphicx}

\newcommand{\msun}{M$_{\odot}$\,}

\def\lesssim{\mathrel{\hbox{\rlap{\hbox{\lower4pt\hbox{$\sim$}}}\hbox{$<$}}}}
\def\gtrsim{\mathrel{\hbox{\rlap{\hbox{\lower4pt\hbox{$\sim$}}}\hbox{$>$}}}}

\def\arcsec{\hbox{$^{\prime\prime}$}}

\def\farcm{\hbox{.\kern -0.7ex\raisebox{.9ex}{\scriptsize$\prime$}}}
\def\farcs{\hbox{\kern 0.13ex.\kern -0.95ex%
\raisebox{.9ex}{\scriptsize$\prime\prime$}\kern -0.1ex}}

\def\it{}

\def\lesssim{\mathrel{\hbox{\rlap{\hbox{\lower4pt\hbox{$\sim$}}}\hbox{$<$}}}}
\def\gtrsim{\mathrel{\hbox{\rlap{\hbox{\lower4pt\hbox{$\sim$}}}\hbox{$>$}}}}

\def\arcsec{\hbox{$^{\prime\prime}$}}

\def\farcm{\hbox{.\kern -0.7ex\raisebox{.9ex}{\scriptsize$\prime$}}}
\def\farcs{\hbox{\kern 0.13ex.\kern -0.95ex%
\raisebox{.9ex}{\scriptsize$\prime\prime$}\kern -0.1ex}}

\def\arcsec{''}

\def\deg{\hbox{$^\circ$}}
\def\la{\mathrel{\hbox{\rlap{\hbox{\lower4pt\hbox{$\sim$}}}\hbox{$<$}}}}
\def\ga{\mathrel{\hbox{\rlap{\hbox{\lower4pt\hbox{$\sim$}}}\hbox{$>$}}}}

\def\arcsec{\hbox{$^{\prime\prime}$}}

\def\farcm{\hbox{$.\mkern-4mu^\prime$}}
\def\farcs{\hbox{$.\!\!^{\prime\prime}$}}

\def\fnum@figure{{\rmfamily Fig.\space\thefigure.---}}%
\def\fnum@table{{\rmfamily Table \thetable:}}%
\def\fnum@plate{{\bfseries Plate \theplate.}}%
\def\fps@figure{bp}%
\def\fps@table{bp}%
\def\fps@plate{bp}%
\def\eps@scaling{1.0}%

\title{Adaptive Optics observations of the Galactic center young stars}

\author{S. Yelda\supit{a}, A. M. Ghez\supit{a},  J. R. Lu\supit{b},
T. Do\supit{c}, L. Meyer\supit{a}, M. R. Morris\supit{a}
\skiplinehalf
\supit{a}UCLA Department of Physics and Astronomy, Los Angeles, CA 90095; \\
\supit{b}Institute for Astronomy, University of Hawaii, Honolulu, HI, 96822; \\
\supit{c}UCI Department of Physics and Astronomy, Irvine, CA 92697
}

\authorinfo{Further author information: (Send correspondence to Sylvana Yelda)\\
E-mail: syelda@astro.ucla.edu}

\begin{document}
\maketitle

\begin{abstract}
Adaptive Optics observations have dramatically improved the quality and 
versatility of high angular resolution measurements of the center of our 
Galaxy. In this paper, we quantify the quality of our Adaptive Optics 
observations and report on the astrometric precision 
for the young stellar population that appears to reside in a stellar disk 
structure in the central parsec. We show that with our improved astrometry
and a 16 year baseline, including 10 years of speckle and 6 years of
laser guide star AO imaging, we reliably detect accelerations in the plane of the sky
as small as 70 $\mu$as yr$^{-2}$ ($\sim$2.5 km s$^{-1}$ yr$^{-1}$) and out to a 
projected radius from the supermassive black hole of 1$\farcs$5 ($\sim$0.06 pc).  
With an increase in sensitivity to accelerations by a factor of $\sim$6 over our
previous efforts, we are able to directly probe the kinematic
structure of the young stellar disk, which appears to have an inner radius of 0$\farcs$8.
We find that candidate disk members are on eccentric orbits, with a mean eccentricity 
of $< e >$ = 0.30 $\pm$ 0.07. Such 
eccentricities cannot be explained by the relaxation of a circular disk with
a normal initial mass function, which suggests the existence of a top-heavy IMF or 
formation in an initially eccentric disk.
\end{abstract}


\section{INTRODUCTION}
\label{sec:intro} 
The proximity of the Galactic center (GC) makes the region uniquely suited
for the study of a supermassive black hole and its effects on the surrounding 
nuclear stellar cluster.  Advancements in technology, in particular adaptive optics 
(AO) systems on large (8-10 m), ground-based telescopes, have afforded us a view
of this dense region at such high angular resolution that individual stellar orbits
of short-period stars can be precisely measured. The orbit of the central arcsecond
star, S0-2, for example, has 
provided the best evidence to date that supermassive black holes (SMBH) exist at the
centers of normal galaxies, and that the black hole at the Galactic center, Sgr A*,
has a mass of $M_{BH}$ = 4.1 $\times$ 10$^6$ \msun and distance of $R_o$ = 8.0 kpc, 
respectively \cite{ghez08,gillessen09}. 

With high precision astrometric and spectroscopic measurements, much can also be 
understood about the nuclear star cluster in the vicinity of the SMBH.  
While the GC is populated predominantly by old, late-type stars ($\sim$1 Gyr), 
spectroscopic observations of the region have
unveiled $\sim$200 massive, young ($\sim$ 6 Myr) stars in the central parsec
\cite{allen90,najarro97,ghez03,paumard06,bartko10} whose origins are 
a puzzle due to the strong tidal forces from the black hole.  
Given the youth of this population and the relaxation timescale in the GC 
($\sim$1 Gyr) \cite{hopman06}, the stars' dynamical properties contain a signature
of their formation history. A particularly prominent feature that has been observed is 
a stellar disk containing a significant fraction (possibly up to 50\%) of the young 
stars orbiting the black hole in a clockwise sense 
\cite{genzel00,levin03,genzel03,paumard06,lu09,bartko09}. The orientation
of the disk orbital plane can be described by its inclination and angle to the 
ascending node, and is roughly $i \sim$ 130$\deg$ and $\Omega \sim$ 100$\deg$,
with a 1$\sigma$ dispersion of $\sim$10$\deg$
\cite{paumard06,lu09,bartko09}. The disk's surface density profile has been 
observed to fall off like $R^{-2}$, which has been the main argument for 
{\em in situ} formation and against cluster infall \cite{paumard06,lu09,bartko09}.

While there is consensus in the literature regarding the existence of the clockwise 
disk and its surface density profile, many of its properties have yet to be 
constrained, in part because interpretations of kinematic studies rely on the ability 
to assign disk membership. The difficulty in determining which stars belong to the 
disk arises from the lack of knowledge about a star's line-of-sight distance relative 
to Sgr A*.  This can be overcome, however, with detections of accelerations in the
plane of the sky \cite{lu09}. While some stars in the central arcsecond ($r <$ 0$\farcs$5)
have been seen to orbit Sgr A* completely \cite{ghez08,gillessen09}, the detection of 
accelerations becomes more difficult at larger radii where the clockwise disk
stars are located ($r >$ 0$\farcs$8) and requires both high precision 
astrometry and patience. In this proceeding, we report on the quality of our
AO observations (\S\ref{sec:obs}), our astrometric precision, including detections
of accelerations in the plane of the sky for the young stars (\S\ref{sec:astrometry}),
and finally the eccentricity distribution of the accelerating disk stars (\S\ref{sec:ecc}).

\section{HIGH ANGULAR RESOLUTION IMAGING}
\label{sec:obs}
The Galactic center region was imaged at high angular resolution with the Keck telescopes 
from 1995-2011. In the first decade, the images were made with the near-infrared 
speckle imaging camera, which provided diffraction limited images 
($\theta$ = 0$\farcs$05) with Strehl ratios of $\sim$0.05. In each final image 
constructed from the speckle data 
($<N_{exp}>$ $\sim$ 1600, $t_{exp} \sim$ 0.14 s, coadd =1), we detect $\sim$160 stars 
on average, down to $K \sim$ 16 in a 5$\farcs$2$\times$5$\farcs$2 field of view. 

Starting in 2004, images were obtained with the Keck II laser guide star adaptive 
optics system in conjunction with the facility near infrared camera, NIRC2 
($<N_{exp}>$ $\sim$ 100, $t_{exp} \sim$ 2.8 s, coadd = 10).  
The tip/tilt star is located at a distance of
$r$ = 19$\arcsec$ from the GC. Despite this relatively large distance,
as compared with speckle imaging these data provide nearly an order of magnitude
improvement in Strehl ratio ($\sim$0.28), a factor of 4 increase in areal coverage 
(field of view of 10$\arcsec\times$10$\arcsec$) and a considerable increase in the average 
number of detected stars ($\sim$2000) down to $K \sim$ 20. Details of these 
observations can be found in
Ghez et al. \cite{ghez08}, Lu et al. \cite{lu09}, and Yelda et al. (in preparation).
With these data, we measure the proper motions and accelerations in the plane of the
sky for a sample of $\sim$70 young stars, which have known line of sight velocities
($v_z$) \cite{paumard06,do09,bartko09} and which are located at projected radii 
$r >$ 0$\farcs$8 - 7$\arcsec$.

\section{HIGH PRECISION ASTROMETRY}
\label{sec:astrometry}
\begin{figure}[!t]
\begin{center}
\includegraphics[scale=0.65]{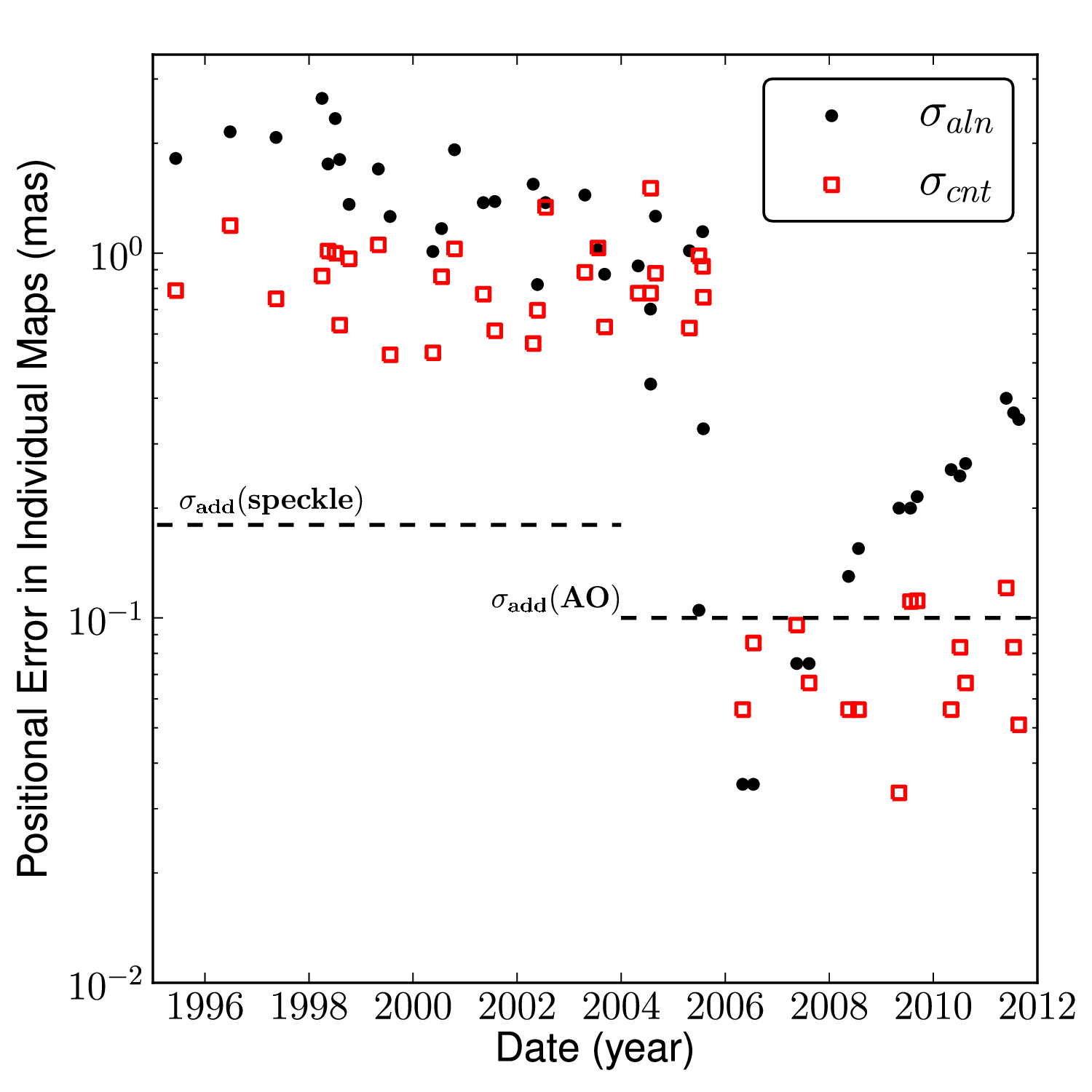}
\caption{Alignment ({\em filled black points}) and centroiding ({\em unfilled red squares}) 
uncertainties as a function of epoch.  The median uncertainty of the young stars
is reported for each epoch. Alignment errors are minimized near the 
reference epoch, 2006 June, and increase with time away from this 
epoch. 
All epochs with $\sigma_{aln} >$ 0.5 mas are from 
speckle imaging, where the higher uncertainties are a result of very few reference 
stars as compared to AO data.  
The additive errors for speckle and AO are shown as dashed lines.
}
\label{fig:posErr}
\end{center}
\end{figure}

There are several sources of statistical uncertainty in the positional measurements
that must be accounted for. While speckle images
have typical centroiding uncertainties of $\sigma_{pos}$ $\sim$ 0.85 mas, the high angular 
resolution AO images allow for nearly an order of magnitude improvement for 
the brightest stars with $\sigma_{pos}$ $\sim$ 0.1 mas ({\em red squares} in 
Figure \ref{fig:posErr}).  This improvement is a result of the higher signal to noise 
ratio in the AO images.
A second source of uncertainty comes from the transformation process that aligns the
stars' positions across all epochs ($\sigma_{aln}$).  These alignment errors are a 
function of time from the reference epoch and of the number of reference stars used 
in the transformation to the so-called ``maser-frame'' in which Sgr A* is at rest
(see Yelda et al. \cite{yelda10} for details of this process).
As seen in Figure \ref{fig:posErr}, $\sigma_{aln}$ is 
minimized near the reference epoch, 2006 June, and is larger for the speckle epochs
($\sigma_{aln} >$ 0.5 mas), which have on average $\sim$6$\times$ fewer reference stars
than are available in AO epochs.
Finally, there is also a source of statistical uncertainty in the positional measurements 
that is not captured in the other terms that 
arises from inaccuracies in the estimates of the PSF wings of neighboring sources.
This additive error term ($\sigma_{add}$) is estimated as the amount of error required
to minimize the residuals between the observed and expected velocity $\chi^2$ distribution
for a given number of degrees of freedom (see Clarkson et al. \cite{clarkson11} and
Yelda et al. (in preparation) for details).  The size of this error term is negligible
for the speckle data ($\sigma_{add}$ = 0.18 mas), but becomes important 
for the AO images ($\sigma_{add}$ = 0.10 mas).

Positions as a function of time are fit for proper motions and accelerations. 
Whether a star exhibits significant accelerated motion in the plane of the sky 
depends on several factors, 
including its distance from the supermassive black hole, the time baseline over 
which it is detected, and the precision with which its positions are measured.
In order to determine whether a star's motion is best described by a velocity
or acceleration model, we first compare the goodness of fit of the two models using
the F-test, in which we estimate the quantity $F = \chi^2_{pm} / \chi^2_{acc}$,
which follows an F-distribution under the null hypothesis \cite{hays}.
Here $\chi^2_{pm}$ and $\chi^2_{acc}$ are the reduced $\chi^2$ values for the  
proper motion and acceleration fits, respectively. 
We require that F is greater than the critical F value at the 4$\sigma$ significance 
level and that a star is detected in more than 30 out of the 45 possible epochs, which 
removes the sources that show unphysical accelerations arising from effects such as 
source confusion or edge effects in speckle images. The resulting six stars 
have reliable radial acceleration measurements at $>$5 sigma. For all other
sources, the proper motion fit is used.

\begin{figure}[!t]
\begin{center}
\includegraphics[scale=0.5]{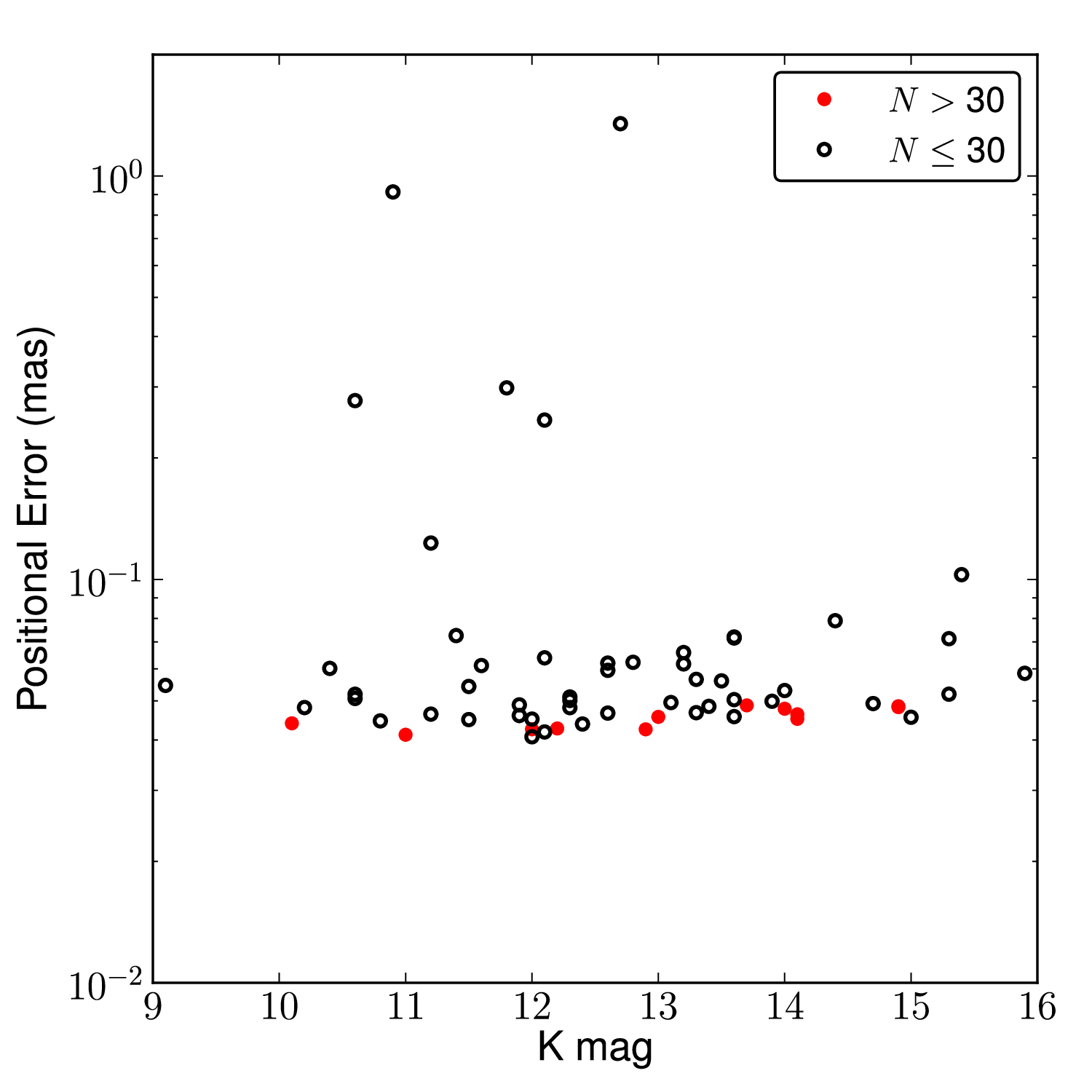}
\includegraphics[scale=0.5]{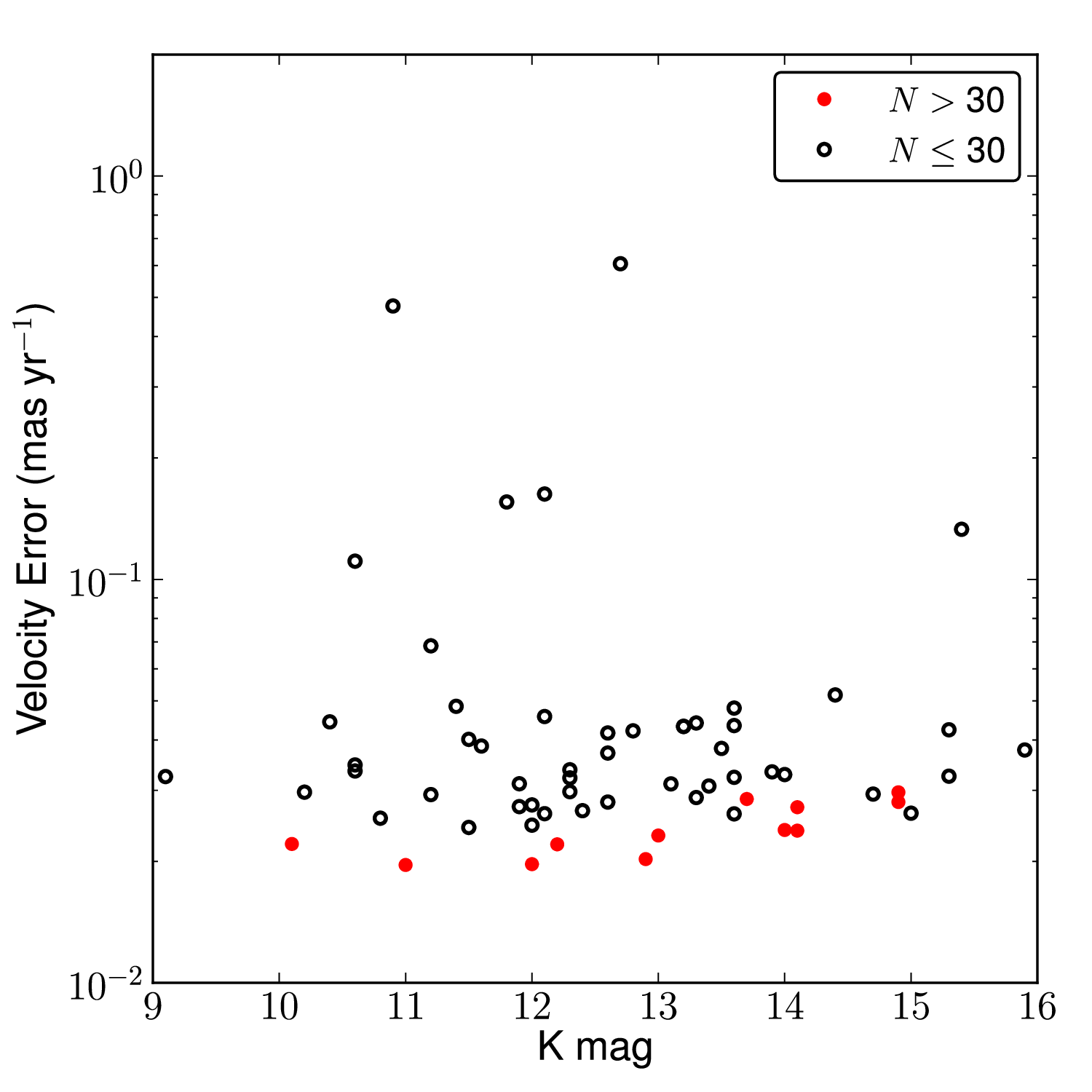}\\
\includegraphics[scale=0.5]{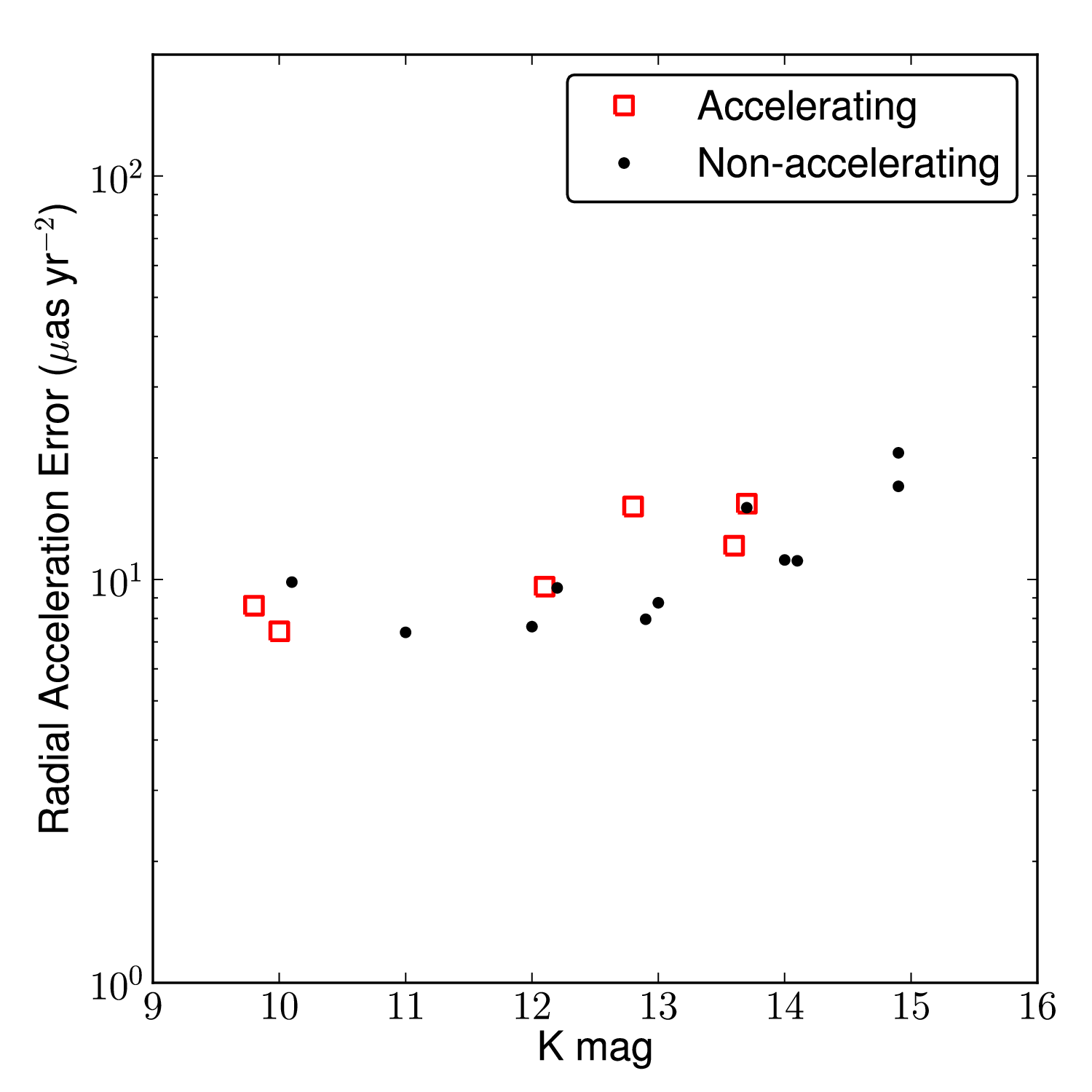}
\caption{Position ({\em top left}), proper motion ({\em top right}), 
and radial acceleration errors ({\em bottom}) as a function
of K magnitude for our sample of young stars beyond a projected radius
of 0$\farcs$8.  Errors are estimated from either the proper motion or 
acceleration fit to each star's individual positions over time.  Stars with
$N >$ 30 epochs ({\em filled red points}) have smaller errors in position
and proper motion than stars in fewer epochs ({\em open circles}). 
The radial acceleration errors are shown for stars in $N >$ 30 epochs, where we
separately plot those six stars passing the F test for accelerations
({\em open red squares}) and those stars for which no detectable accelerations
are found ({\em filled black points}), requiring a non-zero line-of-sight
distance. The average acceleration uncertainty among these stars
is 10 $\mu$as yr$^{-2}$ ($\sim$0.4 km s$^{-1}$ yr$^{-1}$). 
}
\label{fig:pvaErr}
\end{center}
\end{figure}

The position, proper motion, and acceleration errors from the fitting procedure 
as a function of K magnitude ($\lambda$ = 2.2 $\mu$m) are shown in Figure 
\ref{fig:pvaErr} for our sample of young stars. The median errors in positions 
and proper motions are 0.05 mas and 0.03 mas yr$^{-1}$, respectively. Stars
detected in relatively more epochs show smaller errors in positions and 
velocities, as seen in the top panels of Figure \ref{fig:pvaErr}.
Acceleration errors are plotted for the six stars 
that have reliable acceleration measurements, as well as 11 stars with 3$\sigma$
upper limits below their theoretical maximum acceleration (see below).
The average acceleration uncertainty among these stars is 10 $\mu$as yr$^{-2}$ 
($\sim$0.4 km s$^{-1}$ yr$^{-1}$), which is a factor of six improvement over
our earlier efforts in Lu et al. \cite{lu09}.

\begin{figure}[!ht]
\begin{center}
\includegraphics[scale=0.65]{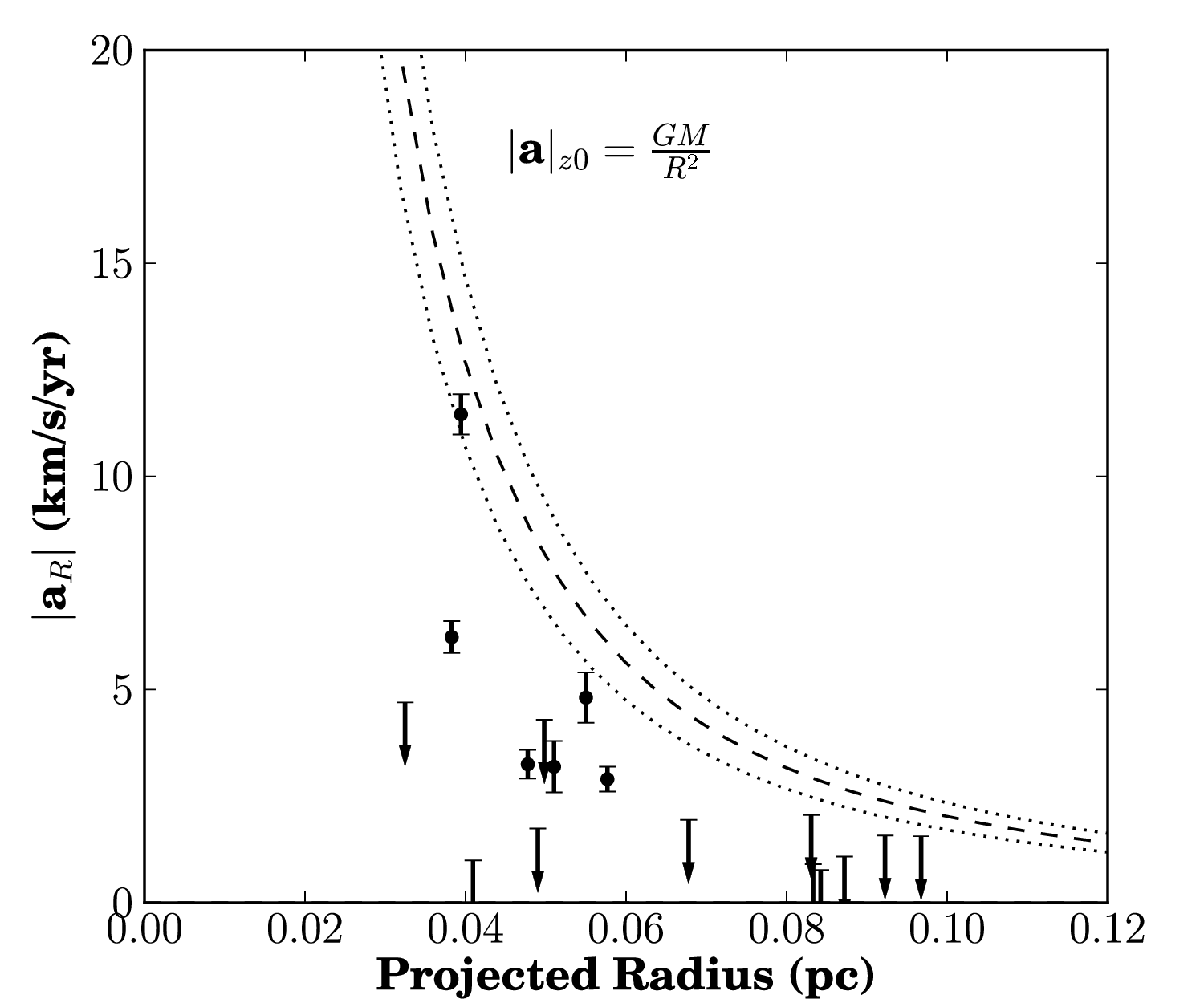}
\caption{Accelerations along the radial coordinate as a function
of projected radius.
The theoretical maximum acceleration ($|a|_{z0}$) for the black hole mass of 
4.6$\times$10$^6$ \msun (see Yelda et al., in preparation) is 
shown as the dashed curve, with the 1$\sigma$ upper and lower boundaries shown
as dotted curves. For our sample of stars at projected radii $r >$ 0$\farcs$8, we 
reliably detect six significant accelerations 
out to R=1$\farcs$5 (0.06 pc), shown with 1$\sigma$ error bars. 
These accelerations yield line-of-sight distances to these sources, allowing for
precise determination of the stars' orbits. Stars with 3$\sigma$ acceleration upper 
limits below the theoretical maximum acceleration (+1$\sigma$) are shown as downward 
pointing arrows and have strong constraints on their line-of-sight distances 
($|z| >$ 0).
}
\label{fig:acc_r2d}
\end{center}
\end{figure}

Figure \ref{fig:acc_r2d} shows the radial acceleration measurements 
as a function of projected radius for stars in the central 10$\arcsec$ field for 
which we have acceleration information. The six significant accelerations are shown 
with 1$\sigma$ error bars. This increases the number
of acceleration measurements beyond 1$\arcsec$ over our previous work in 
Lu et al. \cite{lu09} by a factor of six, or equivalently, an additional five stars,
three of which are reported by Gillessen et al. \cite{gillessen09}. The most distant 
star from the SMBH for 
which an acceleration measurement is made is IRS 16SW, located at 
$R$ = 1$\farcs$5 ($\sim$0.06 pc), which is well outside the inner edge of the 
stellar disk. An additional 11 stars 
have 3$\sigma$ upper limits below the theoretical maximum acceleration given by
each star's projected radius, thereby requiring a non-zero line-of-sight distance.

\section{ORBITS OF THE GALACTIC CENTER YOUNG STARS}
\label{sec:ecc}
With six kinematic variables measured 
($x_0$, $y_0$, $v_x$, $v_y$, $v_z$, $a_R$),
the orbital elements can be estimated if the black hole properties are known.
We use updated SMBH properties, which are described in detail in Yelda et al.
(in preparation). We estimate uncertainties in the orbital period ($P$),
eccentricity ($e$), inclination ($i$), angle to the ascending node ($\Omega$), 
longitude of periapse ($\omega$), and time of periapase passage ($T_0$) by
carrying out a Monte Carlo simulation in which we create 10$^5$ artificial 
data sets. In each data set, we sample the six 
kinematic measurements, as well as the
gravitational potential parameters (the black hole's mass, distance, position, 
and an extended mass component).  For stars without $a_R$, we assume a uniform
acceleration prior, with the minimum acceleration set by assuming the stars are
bound and the maximum acceleration set by the stars' projected radii.
Radial velocities, $v_z$, are measured with 
the Keck AO-assisted integral field spectrograph, OSIRIS \cite{do09}, or taken
from the literature if needed \cite{paumard06,bartko09}.
The kinematic variables are transformed to six orbital elements using the 
analytic orbit equations presented in Lu et al. \cite{lu09}.

\begin{figure}[!ht]
\begin{center}
\includegraphics[scale=0.65]{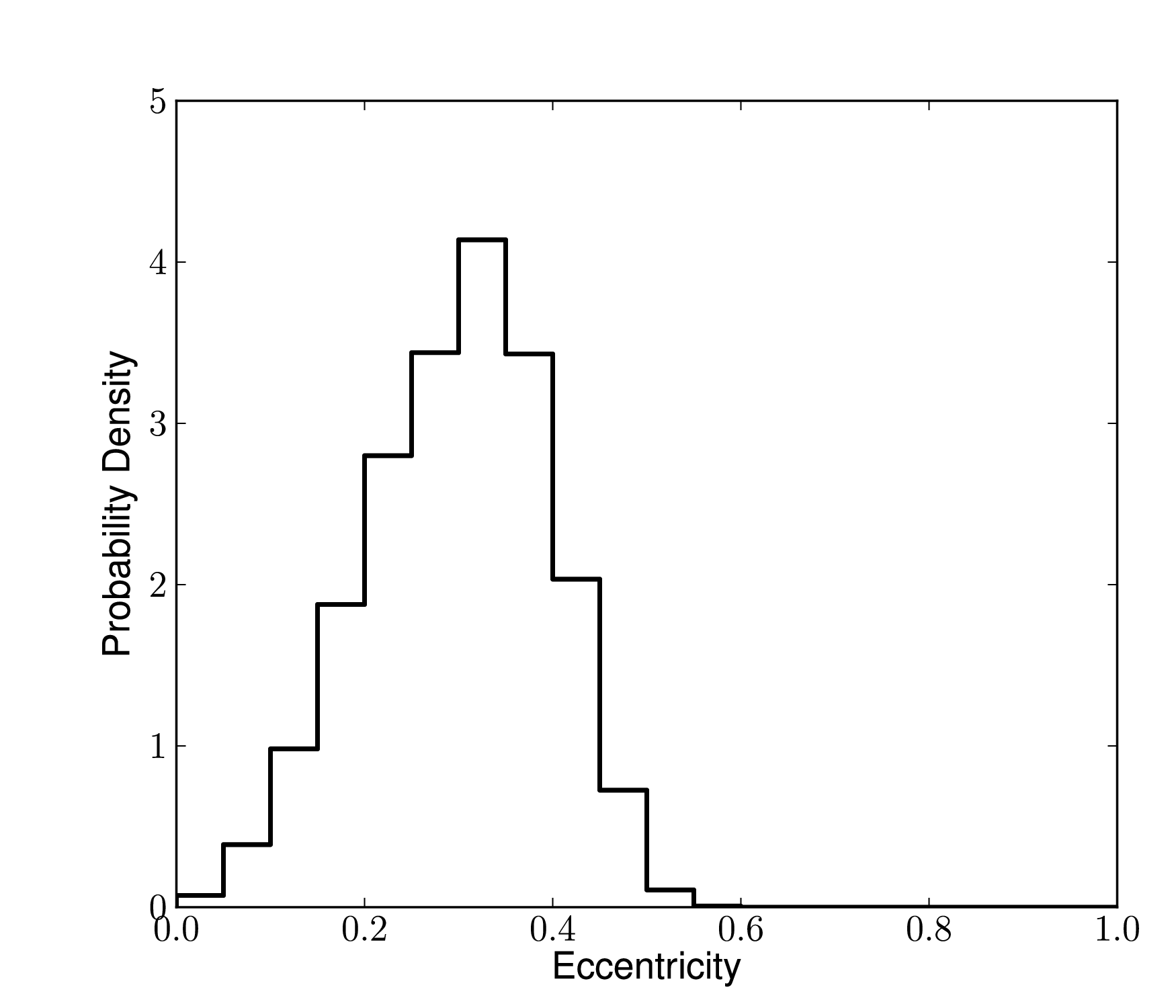}
\caption{Eccentricity distribution from 10$^5$ Monte Carlo trials for the six
accelerating stars. Solutions that are within 15$\deg$ of the clockwise disk
orientation are included. All six stars have orbital planes that are consistent 
with the clockwise disk. These candidate disk stars have eccentric orbits, with 
an average eccentricity of 0.30 $\pm$ 0.07.
}
\label{fig:eccPDF}
\end{center}
\end{figure}

For the six accelerating stars, their line-of-sight distances are determined to within
$\sim$0.01 pc (1$\sigma$) and therefore result in precisely determined orbital parameter
estimates.  Each of these sources has an inclination, $i$, and angle to the ascending
node, $\Omega$, that is consistent with the orientation of the clockwise disk
\cite{paumard06,lu09,bartko09}, making all six stars candidate disk members 
(Yelda et al., in preparation). Figure \ref{fig:eccPDF} shows the combined eccentricity 
distribution for these sources, where only disk solutions (i.e., those solutions that 
are within 15$\deg$ from the orientation of the disk) are included. The orbits for 
these stars have a mean eccentricity of $< e >$ = 0.30 $\pm$ 0.07.

\section{CONCLUSIONS}
Adaptive optics observations have provided a substantial leap forward in the
quality of Galactic center measurements.  As compared to speckle imaging observations,
nearly an order of magnitude improvement is seen in Strehl ratio, number of stars
detected, and astrometric precision.  Such high astrometric precision of stars in
the GC has led to detections of accelerations in the plane of the sky, 
which thereby give a measurement of line-of-sight distances from the black
hole. We present the accelerations and eccentricity distribution for six young stars,
which are all candidate members of the clockwise stellar disk in the central parsec.
The mean eccentricity of the disk stars is $< e >$ = 0.30 $\pm$ 0.07. Given the age
of these stars, self-relaxation in an initially circular disk with a normal IMF cannot 
excite the orbital eccentricities to such high values. Alexander et al. \cite{alexander07}
showed that such eccentricities can be reached only if the IMF were extremely top-heavy.
Although some authors have claimed evidence for a top-heavy IMF 
\cite{nayakshin05,bartko10}, the matter is still under debate (Lu et al., in preparation).
Alternatively, the stars could have formed in an originally eccentric disk.

Astrometric measurements of the Galactic center are currently limited by 
knowledge of the point spread function, which has been assumed to be constant across
the image.  However, both instrumental and anisoplanatic effects 
lead to a spatially- and temporally-variable PSF. Our group is
currently developing a methodology to accurately model both the static,
instrumental PSF (see Fitzgerald et al., this volume) and the time-variable PSF 
using atmospheric profiler data taken simultaneously with our science observations.
Accurate PSF modeling will allow for an improved optical distortion
model and a more stable astrometric reference frame, which is
critical for detecting the effects of general relativity and extended mass on
the orbits of short-period stars at the GC \cite{yelda10}.

\bibliography{gc_bib}

\begin{thebibliography}{10}

\bibitem{ghez08}
{Ghez}, A.~M., {Salim}, S., {Weinberg}, N.~N., {Lu}, J.~R., {Do}, T., {Dunn},
  J.~K., {Matthews}, K., {Morris}, M.~R., {Yelda}, S., {Becklin}, E.~E.,
  {Kremenek}, T., {Milosavljevic}, M., and {Naiman}, J., ``{Measuring Distance
  and Properties of the Milky Way's Central Supermassive Black Hole with
  Stellar Orbits},'' {\em \apj}~{\bf 689},  1044--1062 (Dec. 2008).

\bibitem{gillessen09}
{Gillessen}, S., {Eisenhauer}, F., {Trippe}, S., {Alexander}, T., {Genzel}, R.,
  {Martins}, F., and {Ott}, T., ``{Monitoring Stellar Orbits Around the Massive
  Black Hole in the Galactic Center},'' {\em \apj}~{\bf 692},  1075--1109 (Feb.
  2009).

\bibitem{allen90}
{Allen}, D.~A., {Hyland}, A.~R., and {Hillier}, D.~J., ``{The source of
  luminosity at the Galactic Centre},'' {\em \mnras}~{\bf 244},  706--713 (June
  1990).

\bibitem{najarro97}
{Najarro}, F., {Krabbe}, A., {Genzel}, R., {Lutz}, D., {Kudritzki}, R.~P., and
  {Hillier}, D.~J., ``{Quantitative spectroscopy of the HeI cluster in the
  Galactic center.},'' {\em \aap}~{\bf 325},  700--708 (Sept. 1997).

\bibitem{ghez03}
{Ghez}, A.~M., {Becklin}, E., {Duchjne}, G., {Hornstein}, S., {Morris}, M.,
  {Salim}, S., and {Tanner}, A., ``{Full Three Dimensional Orbits For Multiple
  Stars on Close Approaches to the Central Supermassive Black Hole},'' {\em
  Astronomische Nachrichten Supplement}~{\bf 324},  527--533 (Sept. 2003).

\bibitem{paumard06}
{Paumard}, T., {Genzel}, R., {Martins}, F., {Nayakshin}, S., {Beloborodov},
  A.~M., {Levin}, Y., {Trippe}, S., {Eisenhauer}, F., {Ott}, T., {Gillessen},
  S., {Abuter}, R., {Cuadra}, J., {Alexander}, T., and {Sternberg}, A., ``{The
  Two Young Star Disks in the Central Parsec of the Galaxy: Properties,
  Dynamics, and Formation},'' {\em \apj}~{\bf 643},  1011--1035 (June 2006).

\bibitem{bartko10}
{Bartko}, H., {Martins}, F., {Trippe}, S., {Fritz}, T.~K., {Genzel}, R., {Ott},
  T., {Eisenhauer}, F., {Gillessen}, S., {Paumard}, T., {Alexander}, T.,
  {Dodds-Eden}, K., {Gerhard}, O., {Levin}, Y., {Mascetti}, L., {Nayakshin},
  S., {Perets}, H.~B., {Perrin}, G., {Pfuhl}, O., {Reid}, M.~J., {Rouan}, D.,
  {Zilka}, M., and {Sternberg}, A., ``{An Extremely Top-Heavy Initial Mass
  Function in the Galactic Center Stellar Disks},'' {\em \apj}~{\bf 708},
  834--840 (Jan. 2010).

\bibitem{hopman06}
{Hopman}, C. and {Alexander}, T., ``{Resonant Relaxation near a Massive Black
  Hole: The Stellar Distribution and Gravitational Wave Sources},'' {\em
  \apj}~{\bf 645},  1152--1163 (July 2006).

\bibitem{genzel00}
{Genzel}, R., {Pichon}, C., {Eckart}, A., {Gerhard}, O.~E., and {Ott}, T.,
  ``{Stellar dynamics in the Galactic Centre: proper motions and anisotropy},''
  {\em \mnras}~{\bf 317},  348--374 (Sept. 2000).

\bibitem{levin03}
{Levin}, Y. and {Beloborodov}, A.~M., ``{Stellar Disk in the Galactic Center: A
  Remnant of a Dense Accretion Disk?},'' {\em \apjl}~{\bf 590},  L33--L36 (June
  2003).

\bibitem{genzel03}
{Genzel}, R., {Sch{\"o}del}, R., {Ott}, T., {Eisenhauer}, F., {Hofmann}, R.,
  {Lehnert}, M., {Eckart}, A., {Alexander}, T., {Sternberg}, A., {Lenzen}, R.,
  {Cl{\'e}net}, Y., {Lacombe}, F., {Rouan}, D., {Renzini}, A., and
  {Tacconi-Garman}, L.~E., ``{The Stellar Cusp around the Supermassive Black
  Hole in the Galactic Center},'' {\em \apj}~{\bf 594},  812--832 (Sept. 2003).

\bibitem{lu09}
{Lu}, J.~R., {Ghez}, A.~M., {Hornstein}, S.~D., {Morris}, M.~R., {Becklin},
  E.~E., and {Matthews}, K., ``{A Disk of Young Stars at the Galactic Center as
  Determined by Individual Stellar Orbits},'' {\em \apj}~{\bf 690},  1463--1487
  (Jan. 2009).

\bibitem{bartko09}
{Bartko}, H., {Martins}, F., {Fritz}, T.~K., {Genzel}, R., {Levin}, Y.,
  {Perets}, H.~B., {Paumard}, T., {Nayakshin}, S., {Gerhard}, O., {Alexander},
  T., {Dodds-Eden}, K., {Eisenhauer}, F., {Gillessen}, S., {Mascetti}, L.,
  {Ott}, T., {Perrin}, G., {Pfuhl}, O., {Reid}, M.~J., {Rouan}, D.,
  {Sternberg}, A., and {Trippe}, S., ``{Evidence for Warped Disks of Young
  Stars in the Galactic Center},'' {\em \apj}~{\bf 697},  1741--1763 (June
  2009).

\bibitem{do09}
{Do}, T., {Ghez}, A.~M., {Morris}, M.~R., {Lu}, J.~R., {Matthews}, K., {Yelda},
  S., and {Larkin}, J., ``{High Angular Resolution Integral-Field Spectroscopy
  of the Galaxy's Nuclear Cluster: A Missing Stellar Cusp?},'' {\em \apj}~{\bf
  703},  1323--1337 (Oct. 2009).

\bibitem{yelda10}
{Yelda}, S., {Lu}, J.~R., {Ghez}, A.~M., {Clarkson}, W., {Anderson}, J., {Do},
  T., and {Matthews}, K., ``{Improving Galactic Center Astrometry by Reducing
  the Effects of Geometric Distortion},'' {\em \apj}~{\bf 725},  331--352 (Dec.
  2010).

\bibitem{clarkson11}
{Clarkson}, W., {Ghez}, A., {Morris}, M., {Lu}, J., {Stolte}, A., {McCrady},
  N., {Do}, T., and {Yelda}, S., ``{Proper motions of the Arches cluster with
  Keck-LGS Adaptive Optics: the first kinematic mass measurement of the
  Arches},'' {\em ArXiv e-prints}  (Dec. 2011).

\bibitem{hays}
{Hays}, W.~L.,  [{\em {Statistics}}{\nolinebreak\hspace{0.1em}]}, Harcourt
  Brace College Publishers (1994).

\bibitem{alexander07}
{Alexander}, R.~D., {Begelman}, M.~C., and {Armitage}, P.~J., ``{Constraints on
  the Stellar Mass Function from Stellar Dynamics at the Galactic Center},''
  {\em \apj}~{\bf 654},  907--914 (Jan. 2007).

\bibitem{nayakshin05}
{Nayakshin}, S. and {Sunyaev}, R., ``{The `missing' young stellar objects in
  the central parsec of the Galaxy: evidence for star formation in a massive
  accretion disc and a top-heavy initial mass function},'' {\em \mnras}~{\bf
  364},  L23--L27 (Nov. 2005).

\end{thebibliography}
\bibliographystyle{spiebib}  

\end{document}